# Attosecond nanoscale near-field sampling

B. Förg[1,2], J. Schötz[1,2], F. Süßmann[1,2], M. Förster[1,3], M. Krüger[1,3], B. Ahn[4,5], K. Wintersperger[1], S. Zherebtsov[1,2], A. Guggenmos[1,2], V. Pervak[2], A. Kessel[1], S. A. Trushin[1], A. M. Azzeer[6], M. I. Stockman[1,7], D. Kim[4,5], F. Krausz[1,2], P. Hommelhoff[1,3], M.F. Kling[1,2]

[1]Max Planck Institute of Quantum Optics, Hans-Kopfermann-Str. 1, D-85748 Garching, Germany
[2]Physics Department, Ludwig-Maximilians-Universität München, Am Coulombwall 1, D-85748 Garching, Germany
[3]Department of Physics, Friedrich-Alexander-Universität Erlangen-Nürnberg, Staudtstraße 1, D-91058 Erlangen, Germany
[4]Physics Department, CASTECH, POSTECH, Pohang, Kyungbuk 790-784, Republic of Korea
[5]Max Planck Center for Attosecond Science, Pohang, 790-784, Republic of Korea
[6]Attosecond Science Laboratory, King-Saud University, Riyadh 11451, Saudi Arabia
[7]Department of Physics and Astronomy, Georgia State University, Atlanta, Georgia 30303, USA

**The promise of ultrafast light-field-driven electronic nanocircuits has stimulated the development of the new research field of attosecond nanophysics[1-7]. An essential prerequisite for advancing this new area is the ability to characterize optical near-fields from light interaction with nanostructures with sub-cycle resolution. Here, we experimentally demonstrate attosecond near-field retrieval with a gold nanotip using streaking spectroscopy. By comparison of the results from gold nanotips to those obtained for a noble gas, the spectral response of the nanotip near field arising from laser excitation can be extracted. Monte-Carlo (MC) trajectory simulations in near fields obtained with the macroscopic Maxwell's equations elucidate the streaking mechanism on the nanoscale.**

Photoemission from solids is one of the most fundamental and long-studied electron phenomena in nature. Related photon-electron interactions form the basis for modern optoelectronics, where light can trigger electron transfer, amplification, and emission; vice versa, electron injection and excitation can result in the emission of light.[8] The decrease of the dimensions and increase in speed of electronic and optoelectronic circuitry is paramount for improving their performance, with switching rates possibly approaching optical frequencies in all-optical wide-bandgap devices.[4,5,9] This motivates the development of metrology permitting direct access to fields localized on the nanoscale with sub-cycle resolution.[7] Sub-cycle resolution is furthermore highly desirable in tracing, among others, optical-field-induced formation and subsequent relaxation of collective electron dynamics, transient changes in the optoelectronic properties of nanostructured materials in strong fields, charge interactions, and screening after photoemission.

Light-induced collective electron dynamics may lead to the formation of highly-localized and strongly-enhanced nano-optical near fields (see e.g. Ref. 12). This phenomenon has e.g. furnished optical near-field microscopy with nanometer-scale spatial resolution by overcoming the diffraction limit.[10,11] However, despite recent progress in the attosecond control of nanoscale photoemission[2,6,12] and femtosecond near-field characterization,[13-15] real-time observation of the formation and sub-cycle temporal evolution of nanoscale optical near fields has remained elusive.

Attosecond near field streaking (ANFS) spectroscopy, proposed in 2007 (Ref. 1) and extensively studied theoretically,[16-21] can provide sub-cycle resolution of optical near-field dynamics in nanostructured materials, but has not yet been implemented experimentally. In ANFS an attosecond extreme ultraviolet (XUV) pulse ejects a photoelectron with a high initial momentum from a nanostructure that is excited with a time-delayed few-cycle laser field. The freed electron is accelerated by the oscillatory and spatially decaying near-field after its birth. Resultant momentum shifts depend on the (experimentally controllable) instant of release into the field, providing direct time-domain access to the localized optical field and the related nanoscale electron dynamics.

Figure 1(a) shows our experimental setup for ANFS spectroscopy. The experimental and theoretical approaches are described in detail in the methods section and supporting information (SI). Briefly, phase-stabilized, few-cycle near-infrared (NIR) pulses with a duration of 4.5 fs (full width at the intensity half maximum) and a central wavelength of 720 nm are focused onto a gold nanotip. The laser field excites collective electron dynamics, which, in turn, results in spatially-varying near fields around the tip. Attosecond XUV pulses (of a duration of 220 as at a carrier photon energy of 95 eV) with adjustable delay release electrons from the sample, which are subsequently accelerated in the near fields. The momentum distribution of the freed electrons is recorded as a function of delay between the XUV pulse and the NIR field by a time-of-flight (TOF) spectrometer. The tip can be replaced by a gas target (Ne), allowing independent characterization of the incident NIR field and the XUV pulse by means of standard attosecond streaking.[22] Figures 1(b) and (c) display calculated fields in detection direction, which can be understood as a superposition of incident and scattered fields, clearly revealing a substantial variation of field strength on the nanometer scale. Figure 1(d) shows the response function for the nanotip apex and shank, yielding a phase shift of the near fields with respect to the laser pulse. The near fields at the apex and shank exhibit opposite phase shifts with respect to the laser field, where the shifts are roughly constant for $\lambda$ between 500 nm and 1000 nm. While the near-fields at the tip apex are enhanced by roughly a factor of 2, they are suppressed at the surface of the nanotip shank to about 0.5 with respect to the incident laser field.

Figure 2(a) shows a typical experimental electron kinetic energy spectrum obtained from the two-color interaction with the tips. The spectrum reveals two major contributions: a low-energy contribution below 20 eV and a broader structure between 50 eV and 93 eV with its maximum around 80 eV. Low-energy electrons are mainly generated by strong-field NIR photoemission with a cutoff energy of 15 eV consistent with the applied intensities of ~$1\times10^{12}$ W/cm$^2$, as has been observed in related strong-field photoemission studies on nanotips.[2,6,23] Photoelectrons with an energy exceeding 50 eV are attributed to the XUV photoemission. Spatial scanning of the nanoscale target in the focal plane supports these two different photoemission schemes. Figure 2(b) shows localization of the strong-field NIR photoemission at the tip apex due to the strong nonlinearity of the process; this is a super-resolution phenomenon, which we used for precise positioning of the tip in the laser focus. In contrast, the XUV photoemission is a linear process, leading to electron emission from the whole illuminated surface. The map in Fig. 2(c) thus represents convolution of the shape of the nanotip shape the XUV beam profile.

Figure 3(a) shows a streaking spectrogram obtained from the Au nanotips. The panel to the right depicts the spectrum at a fixed delay. The high-energy edge of the spectrum is assigned to the photoemission of Au-5d electrons. Figure 3(b) shows the streaking spectrogram obtained from Ne for the same pulse parameters. Under the assumption that quantum effects can be neglected and negligible delay for the absolute photoemission time from Ne atoms, the gas streaking gives access to the vector potential of the incident laser field.[24] The extracted streaking curves from Fig. 3(a) and (b) are compared in Fig. 3(c). A negative shift between the streaking traces is directly discernible and is evaluated from this particular set of measurements as $\Delta t = (200 \pm 50)$as.

We performed MC simulations to investigate the origin of the shift. In line with previous theoretical studies[16-19,21] the photoemission process has been assumed to be instantaneous. The retrieved streaking curve from the simulated streaking trace (see SI) from the gold nanotip is shown in Fig. 3(d) (symbols). The curve exhibits a shift with respect to the reference streaking curve calculated using the vector potential of the incident laser (green line) of around $\Delta t = 300$ as with a relative amplitude of about 0.4. We compare the streaking data obtained from the full simulations to curves for electrons emitted from different positions on the tip. The curves obtained for emission from the shank (purple and light blue lines) are in excellent agreement with the retrieved data from the full simulation. For our experimental parameters the emission area on the shank is much larger than at the apex (blue dash-dotted curve in Fig. 3(d)), such that the streaking spectrogram in Fig. 3(a) is dominated by the shank contribution.

So far, no other experimental studies have been reported, which compare streaking delays between a solid and a noble gas. In general the photoemission process is not instantaneous, but electrons are released into the external streaking fields with some effective absolute photoemission time delay  Theoretical studies for neon[25] suggest that the absolute photoemission delay is below 10 as. Theoretical studies for a solid surface, taking into account our emission geometry, are lacking. In our case, the detected electrons from the nanotip shank effectively probe the electric field component parallel to the surface, which is continuous across the surface and approximately homogeneous over the electron emission depth. Therefore, delays due to the transport of the electrons to the surface, which were highlighted in previous studies[26], are largely absent. At the relatively high XUV energies employed in our experiments additional effects should play a minor role (discussion see SI). We therefore assume that the photoemission delay from the nanotip shank is negligible compared to the measured time shift between the streaking spectrograms for the nanotip and gas. The shift is thus attributed to the difference in the electric fields acting on the released electrons, which can be related to the collective free-electron polarization response of the gold nanotip (see SI).

In conventional attosecond streaking spectroscopy from atoms[22,27,28] or solids[29,30], the streaking laser field is homogeneous and the electron momentum change is proportional to the vector potential of the streaking field. In contrast, for inhomogeneous near fields, theoretical studies, carried out for isolated nanoparticles[1,17-20] and nanostructured systems[1,16], have revealed that in this case three different regimes can be identified. They are characterized by the adiabaticity parameter $\delta$ defined as the ratio of the time it takes an electron to leave the near field to the period of the laser pulse. The three categories include the limiting cases of (i) $\delta \ll 1$, defining the "instantaneous" regime, where the near field is probed directly, (ii) $\delta \gg 1$, which we refer to as the "ponderomotive" regime [here, analogous to conventional streaking, the vector potential A(t) of the near field (in the Coulomb gauge: $E(t) = -\partial A(t)/\partial t$) is directly accessed], and the intermediate regime (iii) characterized by $\delta \sim 1$. For the parameters of the experiment, electrons take at least 10 fs to leave the near-field, which is long compared to the optical period of 2.5 fs and the duration of the NIR few-cycle pulse of 5 fs. General considerations (see SI) based on the adiabaticity parameter suggest that the streaking curve should only be shifted on the order of -10 as with respect to the vector potential of the near fields at the emission point. This is confirmed by simulations, revealing a minor contribution of -20 as (Fig. 3(d)), which is small compared to the measured shift. We therefore conclude that electrons are ponderomotively accelerated. The measured streaking curve thus provides direct access to the temporal evolution of the vector potential and consequently also the electric near fields around the nanotip. Figure 4(a) shows the reconstructed electric fields from the measurements on the nanotip shown in Figs. 3(b), see methods and SI for details. Our measurements do not only allow to directly retrieve the near-field but also allow to reconstruct the spectrally-resolved response function for the spectral range of the incident laser pulse at the nanotip shank surface. This is particularly useful since in principle this allows the prediction of the near field for any synthesized light field within the same spectral range.

Figures 4(b) and (c) show the response function in terms of the wavelength-resolved phase shifts and relative amplitudes, respectively, between the local near field and the incident laser field under varying experimental conditions (different days, different tips). The crosses represent data points, whereas the dots show the average response and the respective standard deviation. The average measured phase shift lies between -0.4 and -0.8 rad which is in good agreement with theoretical values of -0.5 to -1.1 (green shaded area), where the spread is caused by variation of the nanotip shape and the simultaneous probing over different emission points. Measured relative amplitudes agree well with the simulations. Presently, the measurements are dominated by the nanotip shank contribution. Increasing the XUV flux and improving the focusing should allow to also decipher contributions from the nanotip apex (see SI).

In conclusion, we have successfully implemented ANFS on Au nanotips. The ponderomotive regime for ANFS provided direct access to local near fields on nanometer spatial and attosecond temporal scales. Our approach is well suited for studying more complex structures, including the characterization of ultrafast nanocircuits. Understanding the local near fields, e.g., in simple optical-field-driven switches, will allow constructing more complex, coupled nanostructures that may lead to the initial building blocks for petahertz electronics. Attosecond nanoscopy can be implemented by scanning the XUV beam (with a smaller focus) over the surface or, alternatively, by using a photoemission electron microscope.

**Figure captions**

**Figure 1. Attosecond near-field streaking.** (a) Experimental setup: few-cycle NIR- and isolated attosecond XUV-pulses with variable delay are focused onto a gold nanotip. High-energy electrons are emitted via XUV-photoionization and subsequently accelerated in the local near-fields. The delay-dependent final kinetic energy is measured using a TOF spectrometer. The nanotip can be replaced by a gas target. (b)-(c) Maximal normalized local field strengths of the component parallel to the cylinder axis (b) along the laser propagation direction and (c) perpendicular to it as obtained from FDTD-simulations. (d) Response function of the $E_y$-component (laser polarization axis) for two representative points at the nanotip apex and shank, blue and purple in (b) and (c), respectively, showing the absolute value (left panel) and phase (right panel). The response of the shank is close to an infinite cylinder calculated using Mie theory (black dashed line). Slight position dependent oscillations occur due to a plasmon launched at the tip apex.

**Figure 2. Position scan with photoelectron spectra from the Au nanotip.** (a) Electron spectrum under combined illumination with XUV- and NIR-light. (b) Integrated electron emission maps from the low-energy region of (a), dominated by strong-field NIR photoemission and (c) for the high-energy region of (a) caused by linear XUV-photoemission. The laser beam propagates in the z-direction. The NIR-photoemission is strongly enhanced at the tip-apex and the XUV-photoemission represents the convolution of the tip geometry with the XUV beam in the focus. The solid black line serves as a guide to the eye showing the outline of the nanotip.

**Figure 3. Analysis of delay shifts between nanotip and gas streaking.** Measured data for (a) the Au nanotip and (b) Ne. The right panels of the spectrograms show electron spectra for a fixed delay of -0.2 fs (nanotip) and 0 fs (gas) illustrating extraction of the streaking curves. A Fermi function (red) is fitted to the cut-off edge of the spectrum. The turning points of the Fermi functions for different delay times provide the curves depicted by symbols in (a) and (b) (details see SI). (c) The retrieved curves are smoothed by Fourier filtering (solid lines) allowing to determine the shift $\Delta t$ between them for every delay. (d) The streaking curve retrieved from a Monte-Carlo simulation (symbols, see SI). The purple and light blue lines illustrate streaking curves for electrons emitted from the front of the shank at $y$=-200 nm and $y$=-3000 nm, respectively. The dashed blue and solid green lines show the streaking curves from the apex and the reference. The inset shows the relation between the simulated streaking curve (solid line) and the local vector potential of the near-field (dashed line) at the emission point.

**Figure 4. Extraction of the electric near-field and response function.** (a) Reconstruction of the local electric near-field (green line) at the nanotip surface from the measured streaking curve (symbols). The error bars indicate 95%-confidence intervals of the Fermi-fit. Retrieved wavelength-resolved (b) relative phase shift and (c) amplitude obtained from different measurements. Data points are shown as crosses. Mean values (symbols) have been obtained by linearly interpolating the retrieved linear response from the individual measurements. Error bars represent the standard deviation. The green shaded areas show the range of phase shifts and relative amplitudes expected from the electric field calculations considering shape variations of the nanotips in the experiments and emission from different points on the surface.

**Figure 1**

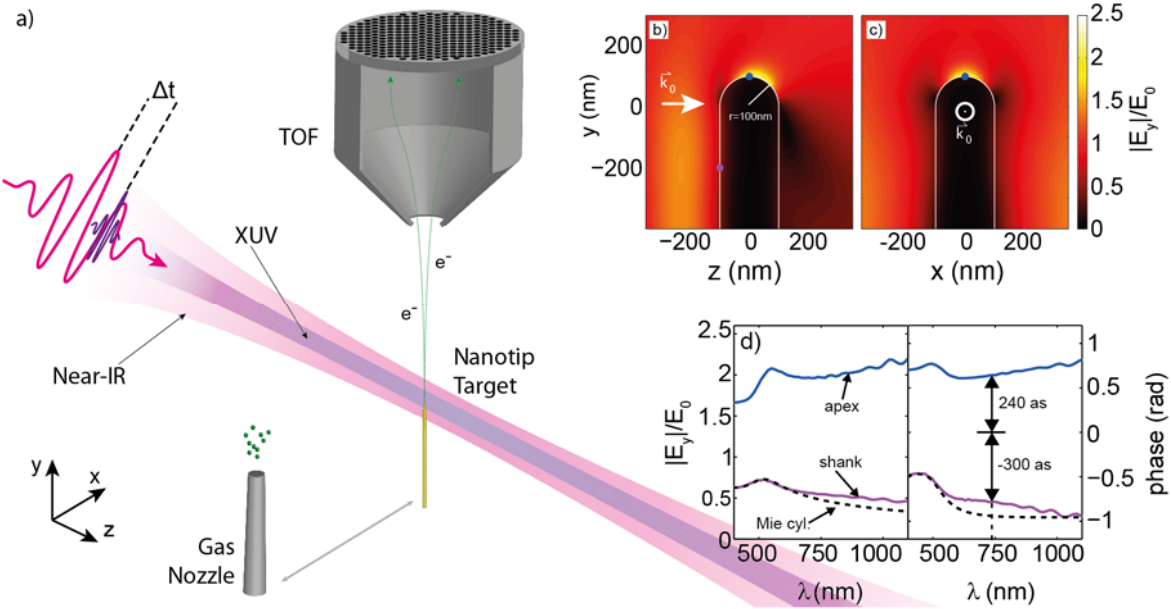

**Figure 2**

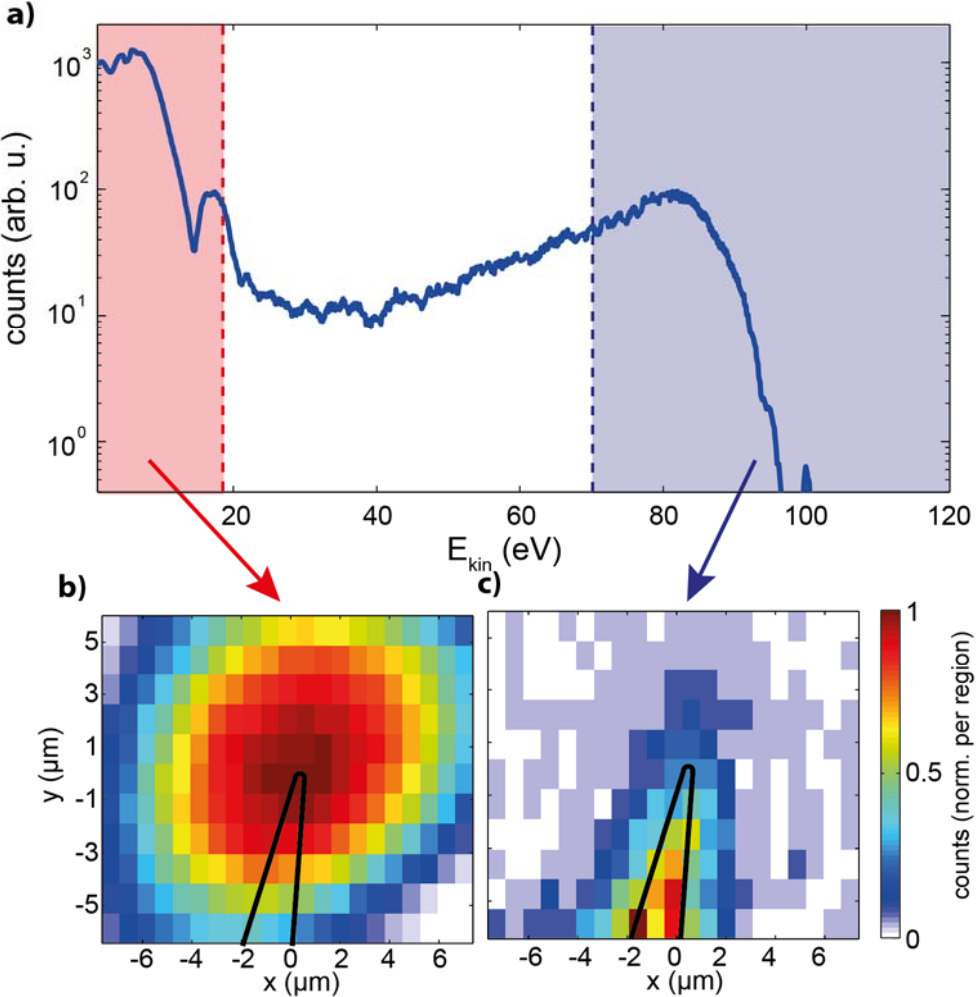

**Figure 3**

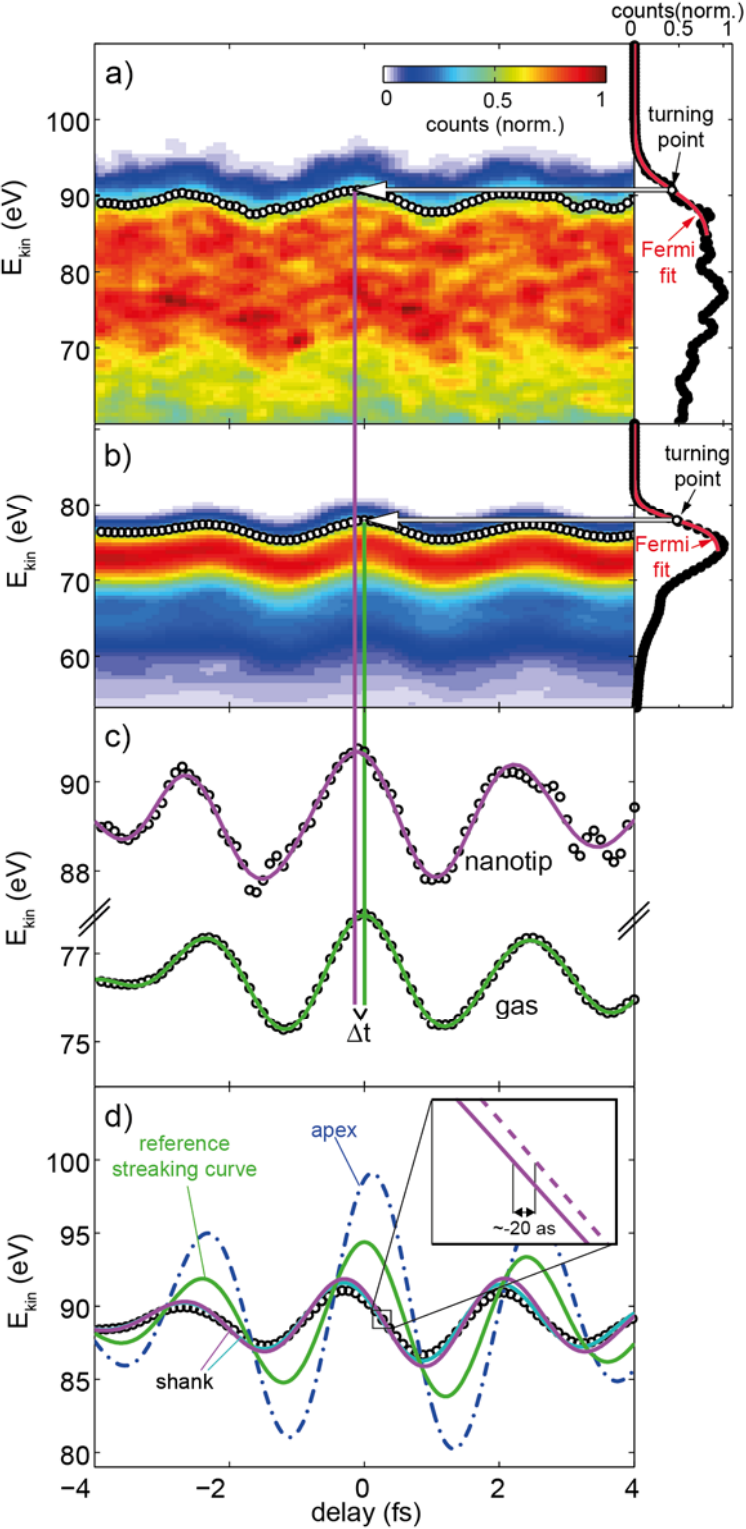

**Figure 4**

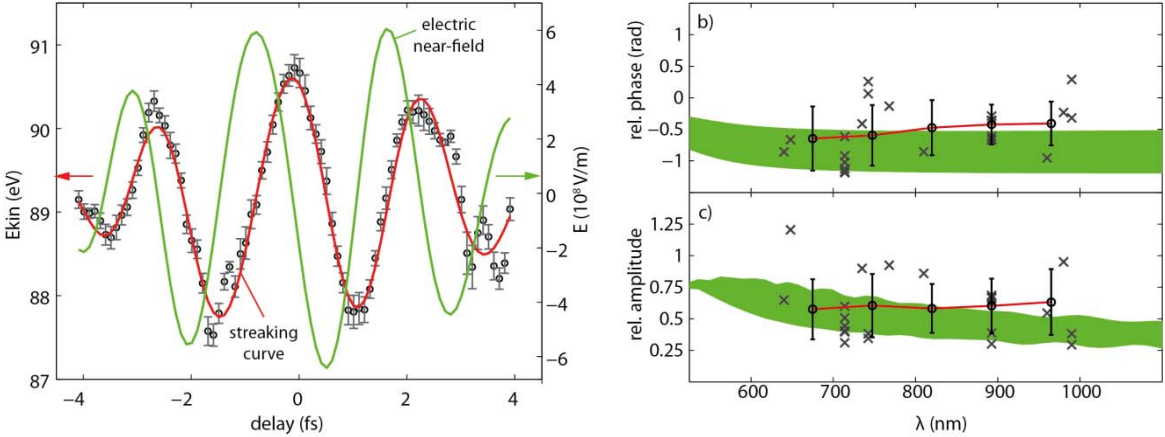

**Methods**

Attosecond near-field streaking:
Few-cycle NIR laser pulses (4.5 fs centered at 720 nm) are produced using a Ti:Sa laser system (Femtopower Compact Pro, Femtolasers) together with a hollow core fiber for spectral broadening and subsequent recompression by chirped mirrors. The vacuum setup used in our experiments is described in more detail in the SI. Briefly, focusing the few-cycle pulses into Ne gas generates high harmonic radiation, while spectral filtering (0.15 $\mu$m thick Zr- foil) provides isolated attosecond pulses (220 as centered at 95 eV). The co-propagating laser pulses are focused (focal length f = 12.5 cm) onto a nanoscopic sample (wet-etched Au-tip, 100 nm tip-radius) using a double-mirror setup, that also introduces a variable attosecond delay between the XUV and NIR pulses.

Three-dimensional motorized stages facilitate positioning the tip-target with nanometer-precision. In addition to the nanotip-target, the stages allow to position a gas nozzle in the laser focus offering the possibility to characterize the NIR-field and the XUV-pulse parameters by recording a reference streaking trace in neon. Photo-excited electrons are propagating through the near-field of the nanotip and the NIR-laser field, the latter polarized parallel to the tip-axis, accelerating or decelerating the electrons on their way to the TOF. The time-of-flight of the electrons is recorded and afterwards converted to kinetic energy. Due to the vanishingly small count rates at high energies of less than 0.1 counts/shot, a lens voltage of 500 V was applied to the TOF to enhance the high energy counts. We carefully checked that the lens has no influence on the measured delays.

Compared to the gas streaking, electron count rates are more than one order of magnitude lower from the nanotip because of the nanoscale size of the sample. This leads to overall acquisition times of up to a few hours in order to obtain reasonable statistics of the streaking trace. The acquisition time is mainly limited by the time the CEP can be locked. To exclude phase drifts and instabilities during measurements, several nanotip and gas streaking spectrograms are recorded, compared and, for stable conditions, superimposed.

In contrast to gas phase streaking, the expanded density of states of the Au nanotip-target leads to severe broadening of the energy spectrum, resulting in streaking curves with a width of more than 20 eV. Inelastic scattering of photoelectrons on their way to the surface additionally broadens the spectra. Irrespective of these effects, analysis of electrons with kinetic energies close to the high energy cut-off of the XUV photoemission provides access to the near-field dynamics at the surface of the Au nanotip.

Nanotip preparation:
Gold nanotips are produced from 0.1 mm thick polycrystalline Au wires in a lamella-drop-off-technique by wet electrochemical etching using 90% saturated KCl as etchant. Using this method tip apex radii between 20 nm to 100 nm can be obtained with an opening angle of typical 10°. Excellent surface quality with roughness less than 0.8 nm can extend up to 300 µm from the apex downwards.30

Simulation details:
The optical near-fields are calculated using the finite-difference time-domain (FDTD) approach for a nanotip with a radius of 100 nm for a Gaussian pulse centered at 720 nm and a duration of 4.5 fs (FWHM). The attosecond near-field streaking process was modeled similar to Ref. 19 assuming an XUV spotsize of 5 μm (FWHM) and an IR-laser intensity of $10^{12}$ W/cm$^2$. For the MC streaking spectrogram, electrons are emitted from the surface taking into account the duration of the attosecond XUV-pulse (220 as FWHM) and with an energy distribution given by the experimental XUV photoelectron spectrum. Subsequently to their photoemission, electrons are propagated classically in the electric near-fields around the nanotip. Only electrons having a final propagation direction within the detection angle of the TOF (45° w.r.t to the tip-axis) are recorded. The streaking curves from selected emission points were calculated assuming a fixed initial energy and emission angle and neglecting the finite duration of the XUV-pulse.

Electric field and response function reconstruction:
In the ponderomotive streaking regime the electric fields can be approximated as homogeneous and the final change of momentum $\Delta p$ of the electrons emitted at time $t_0$ can directly be related to the component of the vector potential $A$ parallel to the emission direction: $\Delta p = -e\, A(t_0)$ (in the Coulomb gauge). This allows a direct reconstruction of the local electric field E(t) from the measured shift $\Delta E_{kin}$ in the kinetic energy of the electrons recorded within the streaking spectrogram:

$$E(t_0) = \frac{1}{2e} \frac{1}{\sqrt{2m_e(E_0 + \Delta E_{kin})}} \frac{\partial \Delta E_{kin}(t_0)}{\partial t_0}$$

where e is the electric charge, $m_e$ the mass of the electron, and $E_0$ is the initial kinetic energy. The values $\Delta E_{kin}$ and $E_0$ are obtained from the extracted streaking curve. The Fourier-filtered curves allow a direct calculation of the derivative and reconstruction of the amplitude and phase of the electric field in a delay dependent manner. The amplitude of the incident laser electric field was corrected for the peak intensity measured via low-energy photoelectron emission from above-threshold ionization (details see SI). The response function is retrieved by Fourier transforming the complex electric fields reconstructed form the gas and nanotip measurements and comparing amplitude and phase in the spectral domain (details see SI).

**Acknowledgement**
We acknowledge the coworkers that have helped to build the attosecond infrastructure, in particular Thorsten Uphues and Adrian Wirth. We are grateful to Ulf Kleineberg for supporting the fabrication of multilayer XUV mirrors and to Seungchul Kim for fruitful discussions. We are grateful for support by the Max Planck Society and the DFG through SPP1391 and the Cluster of Excellence: Munich Centre for Advanced Photonics (MAP). F.S., S.Z., and M.F.K. acknowledge support from the EU via the ERC grant ATTOCO, M.K., M.F. and P.H. via the ERC grant NearFieldAtto. This research has also been supported in part by the Global Research Laboratory program [Grant No 2009-00439], by the Leading Foreign Research Institute Recruitment program [Grant No 2010-00471] and by the Max Planck POSTECH/KOREA Research Initiative [Grant No 2011-0031558] through the NRF. For M.I.S. research, the main support came from Grant No. DE-FG02-11ER46789 from the Materials Sciences and Engineering Division of the Office of the Basic Energy Sciences, Office of

Science, U.S. Department of Energy, and an additional support was provided by Grant No. DE-FG02-01ER15213 from the Chemical Sciences, Biosciences and Geosciences Division, of the Office of the Basic Energy Sciences, Office of Science, U.S. Department of Energy.

**Author information**
B.F., J.S., and F.S. contributed equally to this work. M.F.K. and P.H. conceived the experiment. B.F., J.S., F.S., K.W., and B.A. performed the measurements. A.K., S.Z., and S.A.T. helped with the laser operation and optimization. M.F. and M.K. prepared and characterized the Au nanotips. A.G. and V.P. prepared specialized optical components. J.S. and F.S. developed the Monte Carlo simulation model and performed the simulations. B.F., J.S., F.S., M.F., A.M.A., D.K., P.H., F.K., and M.F.K. evaluated, analyzed and interpreted the results. All authors discussed the results and contributed to the final manuscript. The authors declare no competing financial interests.
Correspondence and requests for materials should be addressed to Matthias F. Kling (matthias.kling@lmu.de) or Peter Hommelhoff (peter.hommelhoff@physik.uni-erlangen.de).